\documentclass{ws-mpla}
\usepackage{graphicx,epsf}
\usepackage[]{latexsym,amsmath,amssymb}
\newcommand{\be}{\begin{eqnarray}}
\newcommand{\ee}{\end{eqnarray}}

\def\theequation{\thesection.\arabic{equation}}
\begin{document}
\markboth{R.~Casadio and L.~Mazzacurati}
{Bulk shape of brane-world black holes}
\title{BULK SHAPE OF BRANE-WORLD BLACK HOLES}
\author{ROBERTO CASADIO}
\address{Dipartimento di Fisica, Universit\`a di
Bologna and I.N.F.N., Sezione di Bologna,
via Irnerio 46, 40126 Bologna, Italy\\
casadio@bo.infn.it}
\author{LORENZO MAZZACURATI}
\address{Dipartimento di Fisica, Universit\`a di
Bologna and I.N.F.N., Sezione di Bologna,
via Irnerio 46, 40126 Bologna, Italy\\
mazzacurati@bo.infn.it}
\maketitle
\pub{Received (Day Month Year)}{Revised (Day Month Year)}
\begin{abstract}
We propose a method to extend into the bulk asymptotically flat
static spherically symmetric brane-world metrics.
We employ the multipole ($1/r$) expansion in order to allow
exact integration of the relevant equations along the (fifth)
extra coordinate and make contact with the parameterized
post-Newtonian formalism.
We apply our method to three families of solutions previously
appeared as candidates of black holes in the brane world and show
that the shape of the horizon is very likely a flat ``pancake''
for astrophysical sources.
\keywords{black holes, extra dimensions}
\end{abstract}
\ccode{PACS Nos.: 04.70.-s, 04.70.Bw, 04.50.+h}
\section{Introduction}
\setcounter{equation}{0}
In recent years models with extra dimensions have been
proposed in which the Standard Model fields are confined to a
four-dimensional world viewed as a (infinitely thin)
hypersurface (the brane) embedded in the higher-dimensional
space-time (the bulk) where (only) gravity can propagate.
Of particular interest is the Randall-Sundrum (RS) model with
one infinitely extended extra dimension ``warped''
by a non-vanishing bulk cosmological constant $\Lambda$
\cite{RS,kaloper}.
In this context, it is natural to study solutions
corresponding to compact sources on the brane, such as
stars and black holes.
This task has proven to be rather complicated and
there is little hope to obtain analytic solutions such as
those found with one dimension less \cite{emparan}.
The present literature does in fact provide solutions on
the brane \cite{maartens,germani,cfm}, perturbative results
over the RS background \cite{katz,tanaka} and numerical
treatments \cite{shiro,wiseman}.
\par
In this letter we investigate how to extend into the bulk
asymptotically flat static spherically symmetric solutions on
the brane.
We hence set the brane cosmological constant to zero by fine
tuning $\Lambda$ to the brane tension, which we denote by
$3\,\sigma$ (in units with $8\,\pi\,G_{(5)}=1$, where
$G_{(5)}$ is the five-dimensional gravitational constant),
i.e.~$\Lambda=-3\,\sigma^2/2$ \cite{RS,shiromizu}.
We then note that the bulk metric in five dimensions
must satisfy ($\mu,\nu=0,\ldots,4$)
\be
^{(5)}R_{\mu\nu}=-\Lambda\,g_{\mu\nu}
\ .
\label{D+1eq}
\ee
On projecting the above equations on the brane and introducing
Gaussian normal coordinates $x^i$ ($i=0,\ldots,3$) and $z$
($z=0$ on the brane), one obtains the constraints
\be
\left.^{(5)}R_{iz}\right|_{z=0}=
\left.^{(4)}R\right|_{z=0}=0
\ ,
\label{Deq}
\ee
where $^{(4)}R$ is the four-dimensional Ricci scalar and use has
been made of the necessary junction equations \cite{israel}.
One can view Eqs.~(\ref{Deq}) as the analogs of the momentum and
Hamiltonian constraints in the Arnowitt-Deser-Misner (ADM)
decomposition of the metric and their role is to select
out admissible field configurations along a given hypersurface
of constant $z$.
Such field configurations will then be ``propagated'' off-brane
by the remaining Einstein equations, namely
\be
^{(5)}R_{ij}=-\Lambda\,g_{ij}
\ .
\label{Ein2}
\ee
The above ``Hamiltonian'' constraint is a weaker requirement than
the purely four-dimensional vacuum equations $R_{ij}=0$ and is
equivalent to $R_{ij}=E_{ij}$, where $E_{ij}$ is (proportional to)
the (traceless) projection of the five-dimensional Weyl tensor on
the brane \cite{shiromizu}.
\par
In the next Section we outline a general procedure to solve the
bulk equations for a given static spherically symmetric brane metric,
which we shall then apply to several cases in Section~\ref{app}.
In Section~\ref{conc} we finally summerize and comment on our
results.
\section{The procedure}
\setcounter{equation}{0}
\label{proc}
We consider five-dimensional metrics of the form
\be
ds^2=-N(r,z)\,dt^2+A(r,z)\,dr^2+R^2(r,z)\,d\Omega^2+dz^2
\label{g}
\ee
where $d\Omega^2=d\theta^2+\sin^2\theta\,d\phi^2$ and $N$,
$A$ and $R$ are functions to be determined.
For such metrics the momentum constraint is identically solved
and the ``Hamiltonian'' constraint reads
\be
2\!\left({N_{\rm B}'\over N_{\rm B}}\right)'
\!\!+\left({N_{\rm B}'\over N_{\rm B}}+{4\over r}\right)\!
\left({N_{\rm B}'\over N_{\rm B}}-{A_{\rm B}'\over A_{\rm B}}\right)
={4\over r^2}(A_{\rm B}-1)
,
\label{H}
\ee
where the subscript B is to remind that all functions are
evaluated on the brane at $z=0$, $\ '\equiv \partial/\partial r$
and we set $R_{\rm B}=r$ thanks to the spherical symmetry
\cite{wald}.
\par
Our strategy to construct the bulk metric is then made of three
steps:
\begin{romanlist}[(ii)]
 \item
choose a metric of the form~(\ref{g}) whose projection on the
brane,
\be
\left.ds\right|_{z=0}^2
=-N_{\rm B}(r)\,dt^2+A_{\rm B}(r)\,dr^2+r^2\,d\Omega^2
\ ,
\ee
solves the constraint Eq.~(\ref{H});
\item
expand such a metric in powers of $1/r$ (four-dimensional
{\em multipole expansion}) to order $n$,
i.e. introduce the quantities
\be
\left.
\begin{array}{c}
N_n(r,z) \\
A_n(r,z) \\
R^2_n(r,z)
\end{array}\right\}
\equiv\sum_{k=0}^n\,{1\over r^k}\,
\left\{
\begin{array}{l}
n_k(z) \\
a_k(z) \\
r^2\,c_k(z)
\end{array}\right.
\ ,
\label{series}
\ee
where $n_k(0)$, $a_k(0)$ and $c_k(0)$ reproduce the particular
solution chosen at step~1 (to order $n$):
\be
\sum_{k=0}^n\,{1\over r^k}\,
\left\{
\begin{array}{l}
n_k(0) \\
a_k(0) \\
r^2\,c_k(0)
\end{array}\right\}
=
\left\{
\begin{array}{c}
N_{\rm B}(r) \\
A_{\rm B}(r) \\
r^2
\end{array}\right\}
+{\mathcal O}\left({1\over r^{n+1}}\right)
\ ;
\label{bou}
\ee
\item
substitute the sum (\ref{series}) into Eq.~(\ref{Ein2}) and
integrate exactly in $z$ the (three) equations thus obtained
for the functions $n_n(z)$, $a_n(z)$ and $c_n(z)$.
\end{romanlist}
\par
This procedure turns out to be particularly convenient for the
problem at hand because it converts Einstein equations
(\ref{Ein2}) into three sets of second order ordinary
differential equations (in the variable $z$) of the form
\be
{{\rm d}^2f_n\over{\rm d}z^2}-\sigma^2\,f_n=F_{k<n}
\ ,
\label{f_eq}
\ee
where $f_n$ is any of the functions $n_n(z)$, $a_n(z)$ and
$c_n(z)$ ($n\ge 1$), and $F_{k<n}(z)$ a functional of the lower
order terms $f_{k<n}$'s and their first and second
derivatives.
The relevant boundary conditions for Eq.~(\ref{f_eq})
are given by the requirement (\ref{bou}) for $f_n(0)$ and the
junction conditions \cite{israel} which imply
${\rm d}f_n/{\rm d}z(0)=-\sigma\,f_n(0)$.
We shall not display the (cumbersome) $F_{k<n}$ for
simplicity, however they turn out to be such that
the Cauchy problem thus defined admits analytic solutions.
The entire procedure can then be executed automatically
with the aid of an algebraic manipulator to determine the
functions $f_n$ recursively, from the lowest order
up~\footnote{For $n=0$ one has a system of three coupled second
order ordinary differential equations for the $f_0$'s.
The corresponding Cauchy problem is solved by the usual warp
factor, $f_0=\exp(-\sigma\,z)$, which is unique as follows
from the usual theorems of uniqueness.
We further note that for $r\to\infty$ our bulk solutions
reproduce the RS space-time \cite{RS} as one expects for
an asymptotically flat brane}, with the only limitation of
the power and memory of the available computer.
\par
Let us now comment on a few more points regarding the above
procedure.
First of all, we wish to stress that there is a large freedom
in the choice of the metric on the brane.
In particular, the coefficients $n_k(0)$ and $a_k(0)$
can be chosen at will, except for the algebraic
constraints following from Eq.~(\ref{H}).
Since such coefficients are related to the shape of the
source, this input represents the physical content of the model.
A second, related point is the convergence of the series
expansion (\ref{series}).
It is in general difficult to pinpoint one parameter (among the
many possible coefficients of the multipole expansion) whose
``smallness'' guarantees that orders higher than $n$ be
negligible.
Because of this, we should consider our results as reliable for
those values of $r$ and $z$ such that
\be
{|f_{n+1}(z)|\over r^{n+1}}\ll
\left|\sum_{k=0}^{n}\,{f_k(z)\over r^k}\right|
\ ,
\label{conv}
\ee
for given values of the parameters $n_k(0)$ and $a_k(0)$.
In general, for a given $z$, such a condition will be
satisfied for sufficiently large $r$.
\section{Application to brane-world black holes}
\setcounter{equation}{0}
\label{app}
As examples of brane metrics, we have considered the solutions
given in Refs.~\cite{maartens,germani,cfm} which can be expressed
in terms of the ADM mass $M$ and the post-Newtonian parameter
(PNP) $\eta$ \cite{will} measured on the brane.
The case with $\eta=0$ (exact Schwarzschild on the brane) is the well
known black string (BS) \cite{chamblin} which extends all along
the extra dimension.
The BS is known to suffer of serious stability
problems~\cite{chamblin,gregory}, e.g.~the Kretschmann scalar,
\be
K^2\equiv\,^{(5)}R_{\mu\nu\lambda\rho}\,^{(5)}R^{\mu\nu\lambda\rho}
={5\over 2}\,\sigma^4
+48\,M^2\,{e^{4\,\sigma\,z}\over r^6}
\ ,
\label{kre}
\ee
diverges on the AdS horizon ($z\to\infty$).
One is therefore led to conclude that black holes on the
brane must depart from Schwarzschild and have $\eta\not=0$.
As was suggested in \cite{cfm}, the interesting cases are those
with $\eta<0$, since $\eta>0$ implies some sort of anti-gravity
effects (see later for further comments).
\par
Short distance tests of Newtonian gravity yield the bound
$\sigma^{-1}<1\,$mm \cite{RS}
and $|\eta|<10^{-3}$ from solar system tests \cite{will}.
Since we want to study astrophysical sources of solar mass size,
in the following we shall often refer to the typical values
\be
M=10^7\,\sigma^{-1}\sim 1\,{\rm km}
\ ,
\ \ \ \
\eta=-10^{-4}
\ .
\label{para}
\ee
In such a range ($M\,\sigma\gg 1$ and $|\eta|\ll 1$)
one finds a qualitatively identical behavior for all brane metrics
in Refs.~\cite{maartens,germani,cfm}, so we shall just give
the results for case~I of Ref.~\cite{cfm} (see also \cite{germani}),
that is
\be
\begin{array}{l}
N_{\rm B}=1-\strut\displaystyle{2\,M\over r}
\\
A_{\rm B}=
{\left(1-\strut\displaystyle{3\,M\over 2\,r}\right)\over
\left(1-\strut\displaystyle{2\,M\over r}\right)\,
\left[1-\strut\displaystyle{3\,M\over 2\,r}\,
\left(1+{4\over 9}\,\,\eta\right)\right]}
\ ,
\end{array}
\label{caseI}
\ee
where $r=r_{\rm h}\equiv 2\,M$ is the event horizon and the
remaining (non-vanishing) PNPs are
$\beta=\gamma=1+{1\over 3}\,\eta$.
\par
We applied the above procedure to the brane metric
(\ref{caseI}) and were able to solve the corresponding
Eqs.~(\ref{f_eq}) up to $n=19$.
For brevity, we just display a few terms:
\be
&&N_5=
e^{-\sigma\,z}\left[
1-{2M\over r}
-{\eta M^2\over 3 r^4}
{\left(1-e^{\sigma\,z}\right)^2\over \sigma^2}\,
\left(1+{M\over r}\right)
\right]
\nonumber
\\
&&A_3=
e^{-\sigma\,z}\left[
1+{2\,M\over 3\,r}\,\left(3+\eta\right)
+{4\,M^2\over r^2}
\left(1+{\eta\over 12}+{\eta^2\over 9}\right)
\right.
\nonumber \\
&&
\left.
\phantom{A_3=e^{-{\sigma\,z}}\left[\right.}
+{8\,M^3\over r^3}
-{2\,\eta\,M\over 3\,r^3}
{\left(1-e^{\sigma\,z}\right)^2\over \sigma^2}
\right]
\label{bulk_g}
\\
&&{R^2_5\over r^2}=
e^{-\sigma\,z}\left[1
+{\eta\,M\over 3\,r^3}
{\left(1-e^{\sigma\,z}\right)^2\over \sigma^2}
\left(1+{2M\over r}+{7M^2\over 2r^2}\right)
\right]
\nonumber
\ee
It is important to note the appearance of positive exponentials
in the metric functions.
Such terms (which also show up at higher orders) are
non-perturbative in $z$, which makes the expansion in $1/r$
preferable (or at least complementary) to the expansion
for small $z$.
\par
\begin{figure}
\centering
\raisebox{4cm}{${R^2_n}$}
\epsfxsize=3in
\epsfbox{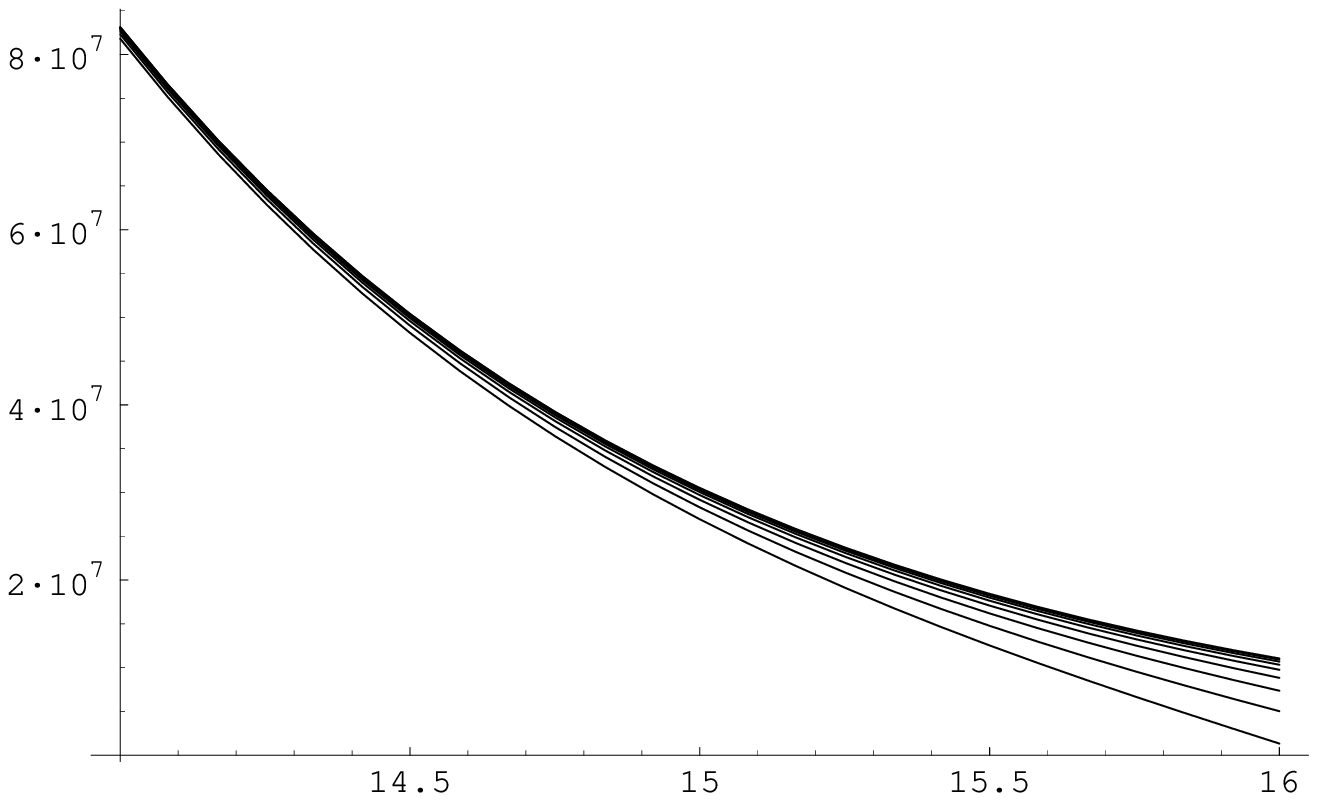}
\\
\raisebox{0.5cm}
{\hspace{5cm}{\bf (a)}\hspace{2.5cm} $z$}
\\
\centering
\raisebox{4cm}{${R^2_n}$}
\epsfxsize=3in
\epsfbox{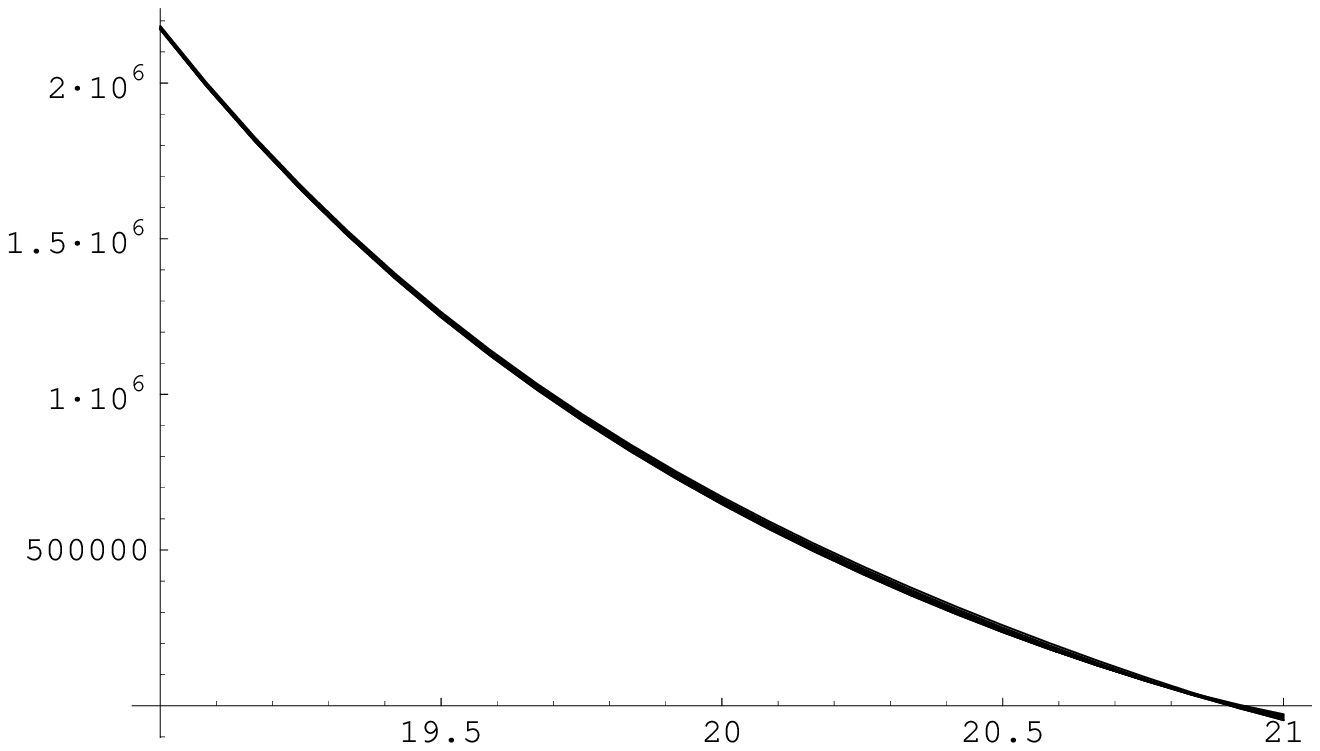}
\\
\raisebox{0.5cm}
{\hspace{5cm}{\bf (b)}\hspace{2.5cm} $z$}
\\
\centering
\raisebox{4cm}{${R^2_n}$}
\epsfxsize=3in
\epsfbox{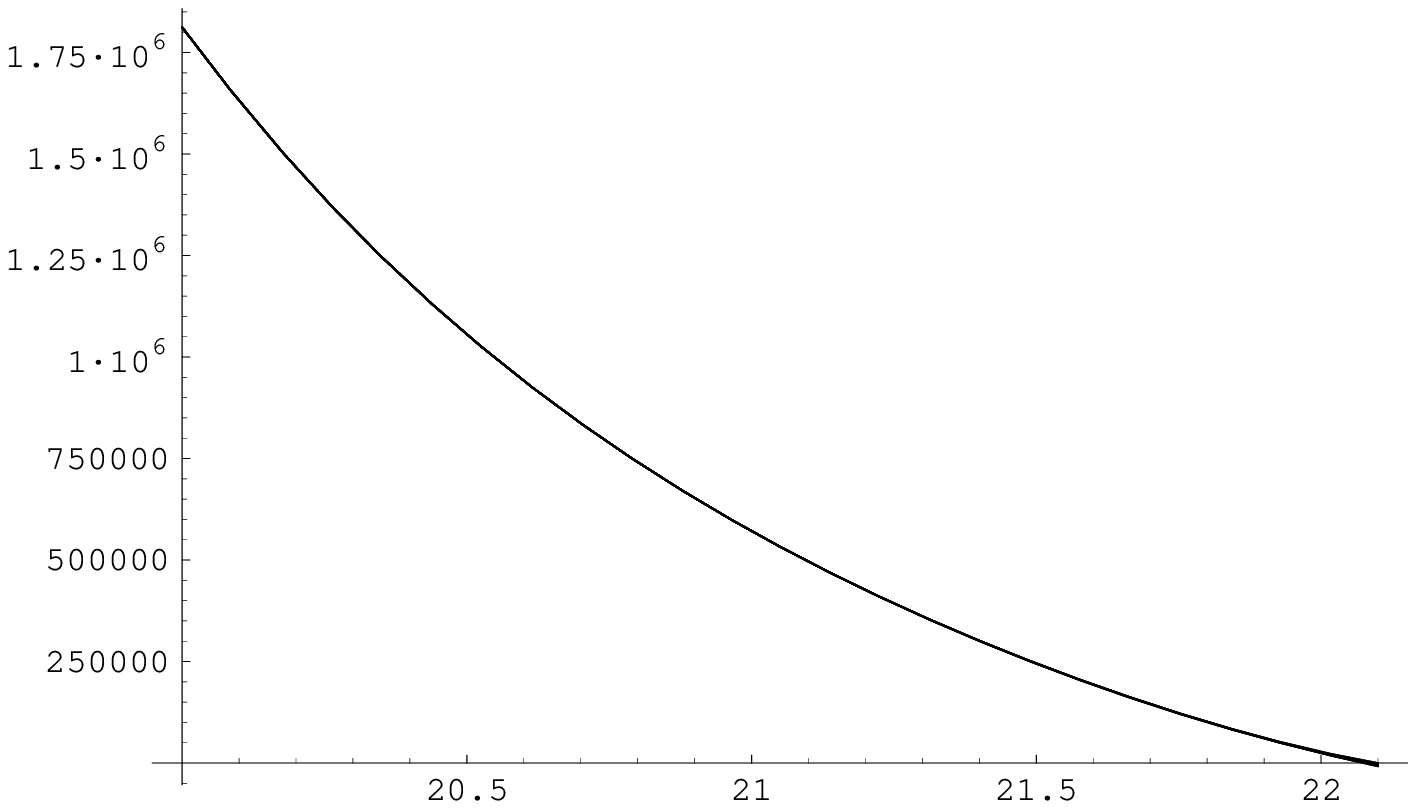}
\\
\raisebox{0.5cm}
{\hspace{5cm}{\bf (c)}\hspace{2.5cm} $z$}
\caption{The function $R^2_n$ near $z^{\rm axis}$ evaluated to
orders $n=10\to 19$ for the data (\ref{para}) and $r=M$ {\bf (a)},
$r=r_{\rm h}=2\,M$ {\bf (b)}, and $r=3\,M$
{\bf (c)} (lengths are in units of $\sigma^{-1}$).
\label{RRrhm}}
\end{figure}
\begin{figure}
\centering
\raisebox{4cm}{${R^2_{19}}$}
\epsfxsize=3in
\epsfbox{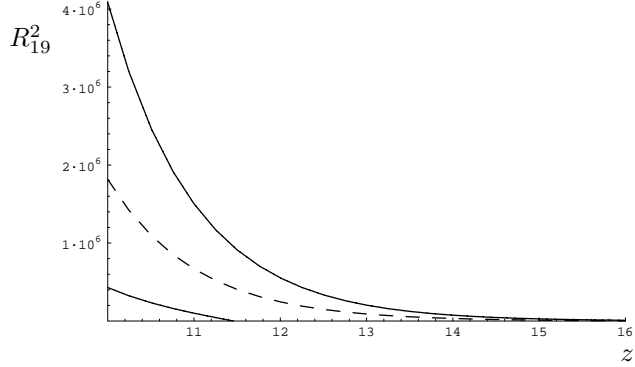}
\\
\raisebox{0.5cm}
{\hspace{8cm} $z$}
\caption{The function $R^2$ evaluated to order $n=19$
for $r=M$, $r_{\rm h}=2\,M$ (dashed line) and $10\,M$
with $M=10^6\,\sigma^{-1}$ and $\eta=-10^{-4}$.
\label{RRrhr}}
\end{figure}

%
%
For $\eta<0$ one generally finds that, for every $n\ge 3$ and
$r>0$, there exists a corresponding value of $z$
[say $z_n^{\rm axis}(r)$] such that $R_n^2(r,z_n^{\rm axis})=0$.
Since $4\,\pi\,R^2$ is the proper area of the sphere
$t=r=z=\,$constant, this seems to indicate that the axis of
cylindrical symmetry is given, in our Gaussian reference frame,
by a line $z=z^{\rm axis}(r)$ (which should be exactly obtained
in the limit $n\to\infty$).
Since $R_n^2$ is determined up to corrections of order
$n+1$, which in general do not vanish for $z=z_n^{\rm axis}(r)$,
one cannot consider this as a mathematically rigorous proof
[the condition (\ref{conv}) obviously fails for $z=z_n^{\rm axis}$].
However, we found that in the physically interesting range of
the parameter space $(M,\eta)$ the $1/r$ expansion yields rather
stable values of $z_n^{\rm axis}(r)$ in a wide span of $n$.
The stability improves for larger values of $r$ [as could
be inferred from (\ref{conv})] and becomes very satisfactory for
$r\gtrsim r_{\rm h}$ (see Fig.~\ref{RRrhm}).
From Eq.~(\ref{bulk_g}) and $M\,\sigma\gg 1$, $r\gg r_{\rm h}$,
one finds
\be
z_5^{\rm axis}\sim{1\over 2\,\sigma}\,
\ln\left({3\,\sigma^2\,r^3\over -\eta\,M}\right)
\ ,
\label{zaxis}
\ee
which numerically agrees fairly well with $z_{19}^{\rm axis}(r)$.
Further, we have checked that $R_n(r_1,z)>R_n(r_2,z)$ for $r_1>r_2$
and $0\le z<z_n^{\rm axis}(r_2)$, so that lines of constant $r$
do not cross and the Gaussian normal form (\ref{g}) of the metric
is preserved for $\eta<0$ in the bulk within our approximation
(for an example see Fig.~\ref{RRrhr}).
\par
\begin{figure}
\centering
\epsfxsize=3.5in
\epsfbox{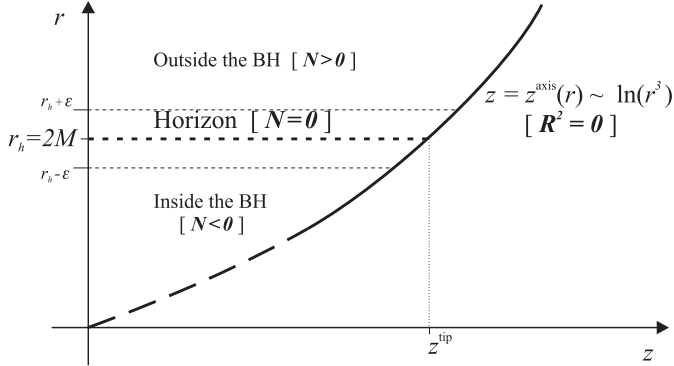}
\caption{Qualitative picture of the bulk structure.
\label{mazza}}
\end{figure}
\begin{figure}
\centering
\raisebox{4cm}{${R^2_{19}\over R^2_{\rm BS}}$}
\epsfxsize=3in
\epsfbox{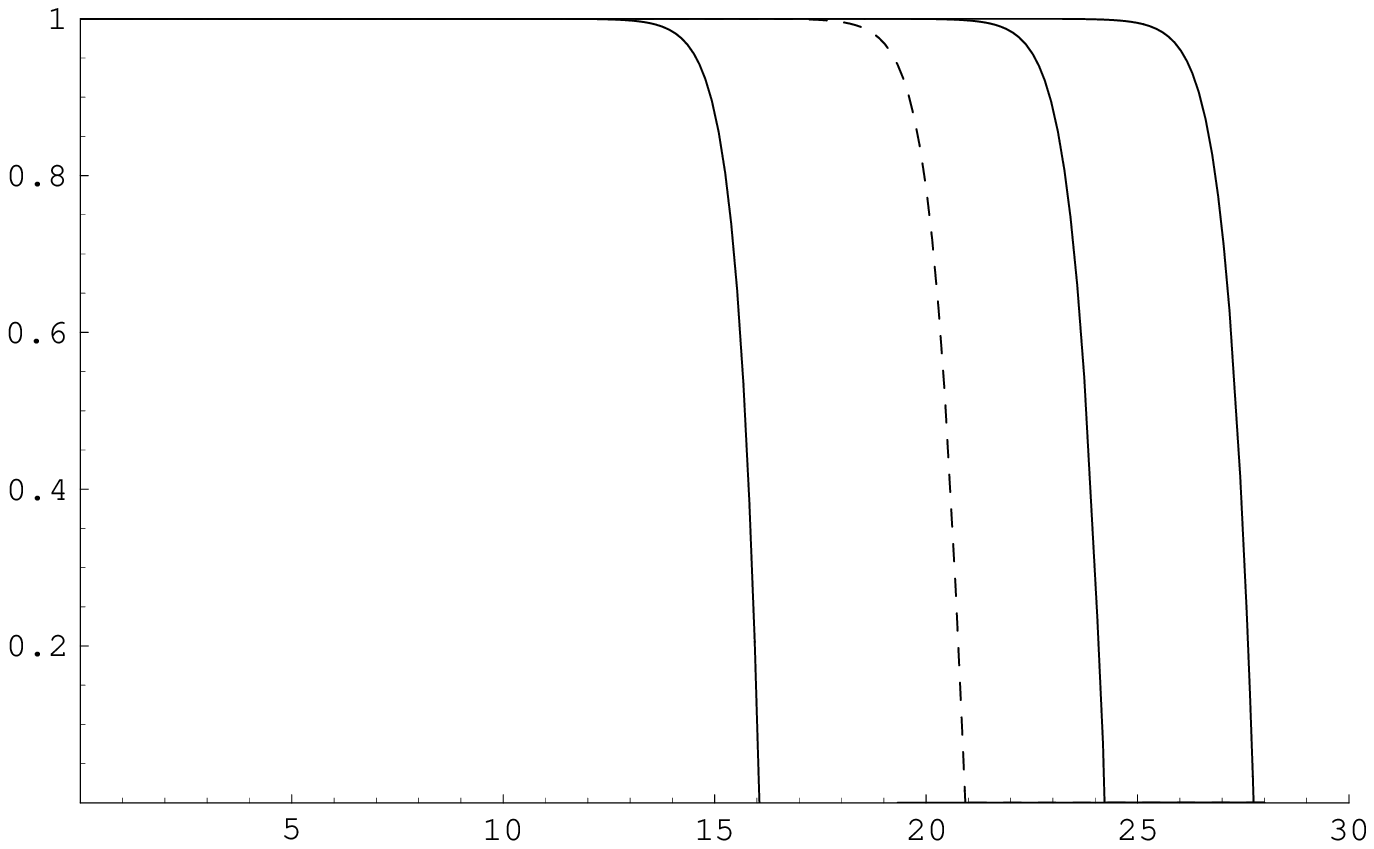}
\\
\raisebox{0.5cm}
{\hspace{8cm} $z$}
\caption{The function $R^2/R^2_{\rm BS}$ evaluated to order $n=19$
for $r=M$, $r_{\rm h}=2\,M$ (dashed line), $10\,M$ and $100\,M$
with  $M=10^7\,\sigma^{-1}$ and $\eta=-10^{-4}$.
\label{RRrh}}
\end{figure}

It is interesting to compare our solutions for $\eta<0$ with the
BS \cite{chamblin}.
In particular, one would like to see the shape of the horizon
in the bulk, knowing that for the BS it does not close
but extends all the way to the AdS horizon.
First we note that, if the horizon closes in the bulk, then
it must cross the axis of cylindrical symmetry at a point
(the ``tip'') of finite coordinates $(r^{\rm tip},z^{\rm tip})$
where $N=R^2=0$.
Of course, we just have such equations explicitly at
order $n$,
\be
N_n(r_n^{\rm tip},z_n^{\rm tip})
=R^2_n(r_n^{\rm tip},z_n^{\rm tip})=0
\ .
\label{tip}
\ee
For large values of $n$, one can solve Eqs.~(\ref{tip})
numerically and find the ``tip''.
More in detail, for $n=19$ one has
\be
N_{19}(r_{\rm h}+\varepsilon,z)>0
\ \ \ {\rm and} \ \
N_{19}(r_{\rm h}-\varepsilon,z)<0
\ ,
\ee
for $\varepsilon\gtrsim (3/100)\,r_{\rm h}$ and
$0<z\lesssim 0.99\,z_{19}^{\rm axis}(r)$.
Thus, to a very good approximation, $N_n(r_{\rm h},z)\simeq 0$
for $0\le z<z_n^{\rm axis}(r_{\rm h})$ when $M\,\sigma\gg 1$
and $|\eta|\ll 1$ and negative.
A good parameterization for the horizon is thus given
by $r\simeq r_{\rm h}$ and
$0\le z\lesssim z_n^{\rm axis}(r_{\rm h})\simeq z_n^{\rm tip}$
(see Fig.~\ref{mazza}).
For the typical parameters (\ref{para}) we have
$r_{19}^{\rm tip}\simeq 0.97\,r_{\rm h}$ and
$z_{19}^{\rm tip}\simeq 20.8\,\sigma^{-1}$
which is very close to
$z_{19}^{\rm axis}(r_{\rm h})=20.9\,\sigma^{-1}$.
This all strongly suggests that the horizon does close in the
bulk, as previously obtained by numerical analysis
\cite{shiro,wiseman} (for a comparison with the BS
see Fig.~\ref{RRrh}).
\par
One can now get an estimate of how flattened the horizon
is towards the brane by comparing the proper length of a
circle on the horizon which lies entirely on the brane,
${\mathcal C}_{||}=2\,\pi\,r_{\rm h}
\simeq 1.3\cdot 10^8\,\sigma^{-1}$,
with the length of an analogous curve perpendicular to
the brane,
${\mathcal C}_{\perp}
\simeq 4\,z_{19}^{\rm tip}\simeq 84\,\sigma^{-1}$.
Since their ratio is huge, one can in fact speak of a
``pancake'' horizon as was suggested, e.g.,
in Ref.~\cite{katz}.
\par
It is interesting to note that for $n=5$, one can still solve
Eqs.~(\ref{tip}) analytically and finds
\be
z_5^{\rm tip}={1\over \sigma}\,\ln
\left[1+{a\,r_{\rm h}\,\sigma\over\sqrt{-\eta}}\right]
\sim{1\over \sigma}\,\ln\left({M\,\sigma\over\sqrt{-\eta}}\right)
\ ,
\label{ztip}
\ee
where $a\simeq 0.5$ and we used $M\,\sigma\gg 1$
in the final expression.
This yields $z_5^{\rm tip}=20.7\,\sigma^{-1}$ for the parameters
(\ref{para}), in excellent agreement with the numerical value
obtained at order $19$.
In light of this stability, one can therefore approximate the
dependence of the exact $z^{\rm tip}$ on the black
hole ADM mass $M$ from Eq.~(\ref{ztip}) and obtains that the
area of the (bulk) horizon is approximately equal
to the four-dimensional (brane) expression~\footnote{The
fundamental (possibly TeV scale) five-dimensional gravitational
coupling $G_{(5)}\sim G_N/\sigma$, where $G_N$ is the
four-dimensional Newton constant \cite{RS}.
Thus, from (\ref{area}) one has
$^{(5)}{\mathcal A}/G_{(5)}\sim M^2/G_N
\sim\,^{(4)}{\mathcal A}/G_N$.},
\be
^{(5)}{\mathcal A}=
4\,\pi\,\int_0^{z_5^{\rm tip}} R^2(r_{\rm h},z)\,dz
\simeq{16\,\pi\over\sigma}\,(2\,M)^2
\ ,
\label{area}
\ee
where we again used $M\,\sigma\gg 1$.
Eqs.~(\ref{ztip}) and (\ref{area}) again supports the idea of a
``pancake'' shape for the horizon (see also \cite{katz} for
the logarithmic dependence of $z^{\rm tip}$ on $M$).
\par
Drawing upon the above picture, in particular the crossing of
lines of constant $r$ with the axis of cylindrical symmetry at
finite $z$, one can infer that the space-times we obtain do not
suffer of one of the instabilities of the BS, namely the diverging
Kretschmann scalar \cite{chamblin}.
In fact, $K^2$ is still an increasing function of $z$ along lines
of constant $r$, but one has
\be
K^2-{5\over 2}\,\sigma^4
=\bar a\,{e^{4\,\sigma\,z}\over r^6}
+{\mathcal O}\left({1\over r^7}\right)
\leq
\bar b\,\sigma^4
+{\mathcal O}\left({1\over r}\right)
\ ,
\ee
where we used Eq.~(\ref{zaxis}) to maximize $K^2$ uniformly
in $z$.
The coefficients $\bar a$ and $\bar b$ depend on the parameters
of the multipole expansion, correctly vanish in pure RS and
$\bar b=48\,M^2$ for $\eta=0$ [the BS, cfr.~Eq.~(\ref{kre})].
The remaining problem of stability under (linear) perturbations
\cite{gregory} is a difficult one to tackle and we do not
attempt at it here.
\par
As we mentioned previously, the cases with $\eta>0$ show a
very different qualitative behavior.
One in fact finds that $R_n^2(r,z)$ is generically a
(monotonically) increasing function of $z$ for all (sufficiently
large) values of $r$, as one would indeed expect on a negative
tension brane \cite{shiromizu}.
However, for any $r_1, r_2>0$ there now exists $z^*_n(r_1,r_2)$
such that $R^2_n(r_1,z^*_n)=R^2_n(r_2,z^*_n)$, i.e.
space-like geodesics of constant $r$ display caustics and the
Gaussian coordinates $(r,z)$ do not cover the whole bulk
\cite{wald}.
\section{Conclusions and outlook}
\setcounter{equation}{0}
\label{conc}
We have explained in some detail a method to extend into the
bulk a given asymptotically flat static spherically symmetric
metric on the brane which is based on the multipole ($1/r$)
expansion.
The application of our method to candidate metrics
\cite{maartens,germani,cfm} for astrophysical sources led
us to conclude that black hole horizon closes in the bulk
and indeed has the shape of a ``pancake''.
Our solutions depend on three parameters
($\sigma$, $M$ and $\eta$), although one could argue that
only two of them are independent.
For instance, in order to recover the four-dimensional
Schwarzschild metric when the extra dimension shrinks to zero
size ($\sigma\to\infty$ in the brane equations \cite{shiromizu}),
one can guess $|\eta|\sim 1/M\,\sigma$.
Reducing the number of dimensions is however
a singular limit of the five-dimensional metric,
so it is critical to obtain precise relations among
the parameters by this procedure.
It is also difficult to extract sensible results for tiny
black holes ($M\,\sigma\ll 1$), e.g. from Eq.~(\ref{ztip})
one now gets $z_5^{\rm tip}\sim M$ and Eq.~(\ref{area})
yields the relation for five-dimensional Schwarzschild
black holes $\,^{(5)}{\mathcal A}\sim M^3$ \cite{myers}.
Our expansion however suggests that the horizon departs
significantly from the line $r=r_{\rm h}$ when
$M\,\sigma\ll 1$ and the above estimate is very rough.
The dependence of the horizon area on the ADM mass is
crucial to study the Hawking evaporation and we hope
to return to it in the future.
\section*{Acknowledgments}
We thank A.~Fabbri for contributing to the early part of the
work.
R.~C.~thanks C.~Germani and R.~Maartens for comments and
suggestions.
\section*{References}
\vspace*{6pt}
\end{document}